\documentclass[showpacs,preprint,aps]{revtex4}
\usepackage{graphicx}
\usepackage{dcolumn}
\usepackage{bm}
\begin{document}
\setcounter{page}{1}
\title
{Time in dissipative tunneling: subtleties and applications}
\author
{N. G. Kelkar$^{a,b}$, D. Lozano G\'omez$^{a}$ and Edgar J. Pati\~no$^a$}
\affiliation{$^a$Departamento de Fisica, Universidad de los Andes,
Cra.1E No.18A-10, Santafe de Bogota, Colombia\\
$^b$Institute of Physics, Jagiellonian University, prof. Stanis{\l}awa 
{\L}ojasiewicza 11, 30-348
Krakow, Poland
}
\begin{abstract}
Characteristic features of tunneling times  
for dissipative tunneling of a particle through a rectangular barrier are studied 
within a semiclassical model involving dissipation in the form of a velocity 
dependent frictional force. 
The average dwell time and traversal time with dissipation are found to be 
less than those without dissipation. This counter-intuitive behaviour 
is reversed if one evaluates the physically relevant 
transmission dwell time.  Apart from these observations, we find 
that the percentage of 
energy lost by the tunneling particle is higher for smaller energies. 
The above observations are tested and 
confirmed in a realistic case by applying the 
dissipation model to study the 
current-voltage data in a Al/Al$_2$O$_3$/Al solid state junction at various 
temperatures. The friction coefficient for Al$_2$O$_3$ as a function of temperature is presented. It is found to decrease with increasing temperature.  
\end{abstract}
\maketitle
\section{A brief history of tunneling time}
Quantum tunneling was one of the first bizarre implications of quantum
mechanics which at first found its application in the study of alpha decay
of radioactive nuclei \cite{gamow}.
It was however soon realized that this phenomenon
was not restricted to nuclear physics but was rather a general result of
quantum mechanics which is now often used in atomic physics \cite{atomic}, 
solid state physics \cite{weAPL}, chaotic scattering \cite{chaos}, 
in constructing electron tunneling
microscopes and even in branches of science other than physics.
Though we are a long way from 1928 when Gamow published his pioneering
work \cite{gamow} and tunneling seems to be a well understood phenomenon
with ramifications in many branches of physics, there still
exist paradoxes and unanswered questions in this field.
For example, the time spent by a particle tunneling a barrier has been a 
topic of much debate with many different definitions of tunneling times 
in literature \cite{mugabooks,haugereview,winreport,buettiker}. In \cite{haugereview}
the authors discussed several time concepts such as the dwell time \cite{smith}, 
traversal time \cite{buetikprl}, phase time \cite{wigner} and 
even complex times. 
Ref. \cite{bertulani} 
discusses the tunneling of composite particles, resonant tunneling and how 
the coupling between intrinsic and external degrees of freedom can affect 
tunneling probabilities. 
The dwell time formalism for the transition
from a quasilevel to a continuum
of states was discussed in the context of electron and alpha particle
tunneling in \cite{price}.
Over the years, many of the time 
concepts have been put to test in physical situations and the transmission dwell time 
seems to emerge as the concept with a physical meaning 
\cite{weAPL,weEPL,wetimedelay} 
as well as free of paradoxes \cite{gotoiwamoto} and singularities such as
those found in the phase time \cite{meprl}. 

Dissipative tunneling times have however been explored to a lesser extent 
historically. In recent years, the authors in \cite{bhataroyAMP} have studied 
the phase and dwell times with dissipation in different contexts \cite{bhataroy2}. 
In \cite{bhataroyAMP}, studying the dissipative tunneling through an inverted 
harmonic oscillator in context with ion transport at nanoscale, the authors 
showed that the phase time delay can be estimated directly in terms of a 
frictional coefficient. The average dwell time, $\tau_D$, through a 
rectangular potential barrier using a path decomposition technique was
investigated in \cite{konno} leading to the counter-intuitive result that 
$\tau_D$ in the presence of dissipation becomes smaller than that in a 
non-dissipative case. The traversal time behaviour for a rectangular barrier 
with energy losses included was described in \cite{ranfagni} 
within a semiclassical approach with 
dissipation included in the form of a frictional force. Using a somewhat 
similar approach for dissipation with the latter given by 
a frictional force as in \cite{buetikprl}, 
in the present work we shall show that the 
counter-intuitive result found in \cite{konno} can indeed be explained.
 
An understanding of the tunneling times with dissipation can prove 
important for studying the characteristics of solid state tunnel junctions. 
The importance of the tunneling times in this context was realized 
in \cite{schnupp} where the author noticed that ``the image force 
acting on an electron tunneling through a dielectric film
enclosed by metal electrodes depends on the dielectric constant 
of the film and the charge build-up in the electrodes which in turn are 
both dependent on the duration of the tunneling process". 
In what follows, we shall present the expressions for the dwell and 
traversal times with dissipation for tunneling through a rectangular barrier. 
Calculations using these expressions are done in context with an 
experiment \cite{weAPL} reported in an earlier work by two authors 
of the present work. 
Ref. \cite{weAPL} presented a method to extract the average dwell times from  
tunneling experiments in solid state junctions. 
The current-voltage (I-V) characteristics reported 
there are now used in a model that includes the effects of dissipation on 
tunneling times. Furthermore, the new fits to these data allow us to 
determine the frictional coefficient for 
Al$_2$O$_3$ from 3.5 to 300 K. 

\section{Semiclassical Dwell and traversal times} 
The concept of an average dwell time was first introduced in the form of a 
collision time by Smith \cite{smith}. Calling it as the time of residence in 
a region and using steady state wave functions he defined it as the integrated 
density divided by the flux in (or out). The lifetime of an unstable state 
was thereby given as the difference between the residence time with and 
without interaction. This difference was essentially the time delay introduced 
due to the formation, propagation and decay of the unstable state. 
Using the residence or dwell time delay, 
he went on further to construct a lifetime matrix, 
{\bf Q}, which 
was Hermitian and the diagonal elements $q_{ii}$ gave the lifetimes 
of unstable states or resonances. The physical relevance of the 
residence time delay (or dwell time delay) 
as well as its relation with the phase time delay introduced earlier by 
Wigner and Eisenbud \cite{wigner} became evident in \cite{wetimedelay} 
and motivated further 
investigations of the same in multichannel scattering \cite{ourpra}. 
The extraction of resonances from multichannel 
scattering data is an involved task and 
we refer the reader to \cite{workman} for details. 
Time delay can also be negative and interesting interpretations of 
time advancements can be found in \cite{timeadv}. 
Many years after Wigner, Smith and Eisenbud's papers, 
the dwell time in one-dimension was revisited by 
B\"uttiker \cite{buettiker} in relation with the newer concepts of 
Larmor time and traversal time. 

Given an arbitrary barrier $V(x)$ in one-dimension, 
for a particle of mass $m$ confined to an interval 
$(x_1, x_2)$, the average dwell time which is defined as the ratio 
of the density to flux can be written within the 
semiclassical Jeffreys-Wentzel-Kramers-Brillouin (JWKB) 
approximation as \cite{weEPL},
\begin{equation}\label{dwellav}
\tau_D (E) \, = \, 
{m \over \hbar}  
\, \int_{x_1}^{x_2} \, {dx \over \kappa(x)} \, {\rm exp} \,\biggl 
[ - 2 \, \int_{x_1}^x \, dx^{\prime} \, \kappa (x^{\prime}) \biggr ] \, , 
\end{equation} 
where, $\kappa(x)\,=\,{\sqrt{ 2\,m\,(V(x)\,-\,E)} / \hbar}$ for tunneling 
at energies, $E$, below the top of the barrier. Defining an effective 
velocity, $v(x) = \hbar \kappa(x)/m$, the traversal time which comes closer 
to the classical definition of a particle ``traversing" a barrier is given 
by, 
\begin{equation}\label{traver}
\tau_{tr} = \int_{x_1}^{x_2} \, {m \over \hbar \kappa(x)}\, dx \,= 
\,\int_{x_1}^{x_2} \, 
\biggl [ {m \over 2(V(x) - E)} \biggr ]^{1/2} \, dx\,.
\end{equation}
\\
{\it Improved JWKB near the base of the barrier}

The JWKB approximation is known to be reasonably good away from the 
lower and upper extremes of the potential barrier. There exist different 
prescriptions to improve the JWKB formulae near the top as well as the 
bottom of the barrier \cite{weWKB}. In the present section, however, we shall 
consider a method to improve the JWKB dwell time only near the base of the 
barrier for reasons which are explained briefly in the paragraph below. 

Dissipative effects in the present work are included by introducing a 
damping force which causes a loss of energy of the tunneling particle.  
In the next section, we will see that the damping force is proportional 
to the velocity of the incoming particle. 
In a rectangular barrier of height $V_0$, the effective velocity, 
$v(x) = \hbar \kappa(x)/m$ becomes,
$v_0 = [2(V_0 - E)/m]^{1/2}$. Thus, inside the tunnel barrier the effective  
velocity decreases with increasing energy. Since it approaches zero near the 
top of the barrier, the energy loss is greater for particles with smaller 
energies relative to the barrier height. 
Therefore, to include dissipative effects, the lowest energies 
(at the base of the barrier) are the most important in the improvement of 
the JWKB formula. With this aim we use the prescription given in [27], 
for the JWKB wave function, in order to improve the dwell time given in 
Eq. (\ref{dwellav}) within the JWKB approximation.

In \cite{eltWKB}, the 
authors used a generalization of the JWKB connection formulas to derive 
expressions of the transition amplitudes which behave correctly at the 
bottom of the barrier. The procedure effectively involved the introduction 
of a multiplicative factor into the normalization of the JWKB wave function. 
Details of the approach with a wide range of examples 
are given in a more recent review article \cite{friedrichrep}. Here, we shall breifly 
explain the derivation of this factor. 
The authors begin by examining
the connection formulas at the classical turning point ($x_t$),
which for example, in the most general case can be written as,
$$ {2 \over \sqrt{\kappa(x)}} \cos{\biggl({1\over \hbar} \biggl | \int_{x_t}^x\, 
\kappa(x^{\prime}) dx^{\prime} \biggr |\, - \, {\phi \over 2} \biggr ) }
\leftrightarrow {N \over \sqrt{\kappa(x)}}\, \exp{\biggl(-{1\over \hbar} \biggl| 
\int_{x_t}^x \kappa(x^{\prime}) dx^{\prime} \biggr | \biggr ) } 
$$
where $\kappa(x) = \sqrt{|2 m (V(x) - E)|}$ and the two parameters $N$ and $\phi$ are
determined by comparing the exact solution corresponding
to an exponentially decreasing wave on the classically forbidden side with
the oscillating JWKB waves on the allowed side.
The connection formulas in conventional
semiclassical theories are derived assuming that the potential is a linear function
of $x$ in the vicinity of $x_t$ and this region extends
``sufficiently far" which eventually leads to $\phi = \pi/2$ and $N = 1$.
Among several examples, the authors consider the case of a rectangular barrier
of height $V_0$
and note that the potential appears to be a
sharp step at each of the two turning points in the rectangular barrier. The
amplitude factor $N$ and phase $\phi$ are obtained by comparing the JWKB waves on
either side of a turning point with the exact wave function for a sharp potential step
of height $V_0$. The result is
$N = 2 \sqrt{{k \kappa_0 / k^2 + \kappa_0^2} } $
with $k = \sqrt{2 m E}$ and $\kappa_0 = \sqrt{2 m (V_0 - E)}$. 
Such an amplitude factor improves the transmission coefficient calculated in 
the JWKB approximation and brings it quite close to the exact one. An example 
of such an improvement can be seen in Fig. 8 of Ref. \cite{friedrichrep}. 

The average dwell time given in Eq. (\ref{dwellav}) is evaluated using the
standard definition \cite{weEPL}, namely,
$\tau_D(E) = \int_{x_1}^{x_2}\, {|\Psi(x)|^2 / j}$
where $j = \hbar \kappa(x)/ m$ (with $\kappa(x) = \sqrt{2 m (V(x) - E)}$)
and $\Psi(x)$ is replaced by the JWKB wave
function $\Psi(x) = \exp{[-2\int_{x_1}^x dx^{\prime} \kappa(x^{\prime})]}$.
If we introduce the improved wave function as discussed above, i.e.,
$\Psi(x) = N \, \exp{[-2\int_{x_1}^x dx^{\prime} \kappa(x^{\prime})]}$, with
$N$ as given above, the average dwell time for a rectangular barrier changes to
$\tau_D^{\rm improved} = N^2 \tau_D$.
The inclusion of dissipation in the present work changes the rectangular barrier
with height $V_0$ to $V(x) = V_0 + \eta v_0 x$. The additional term with $\eta$
is a small perturbation to the rectangular barrier and hence we assume the
factor $N$ for the barrier $V(x)$
to be the same as that in the case of a rectangular barrier and evaluate
$\tau_D^{\rm improved}$ as above.
In Fig. 1, we see a comparison 
between the average dwell times $\tau_D$ without dissipation, 
evaluated exactly for a rectangular barrier \cite{buettiker} (solid line), 
in the JWKB (dashed line) and the improved JWKB (dash dotted line) 
which almost coincides with the exact expression for all energies except ones 
close to the top of the barrier. Notice that in contrast to the exact 
$\tau_D$, one  evaluated in the JWKB does not 
approach zero for energies at the bottom of the barrier. The improved 
JWKB expression takes care of this deficiency and agrees with the exact $\tau_D$. 
\begin{figure}[ht]
\includegraphics[width=9cm,height=9cm]{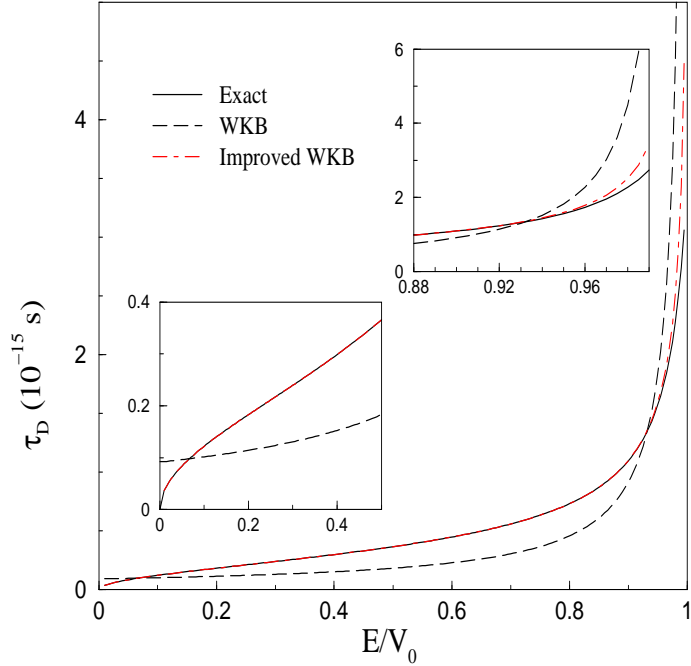}
\caption{\label{fig:eps1}
Average dwell time, $\tau_D$, without dissipation for a rectangular barrier of 
height $V_0$. Solid line is the exact expression as in \cite{buettiker}, 
dashed line is $\tau_D$ within the JWKB approximation and dash-dotted line is 
$\tau_D^{\rm improved}$ (see text).
}
\end{figure}
   
\section{Dissipation in tunneling} 
The question, ``what is the effect of dissipation on tunneling?", was put 
forth in a seminal paper by Caldeira and Leggett \cite{caldeira} in 1983 where 
the authors started with  
a damped equation of motion for the system as follows:
\begin{equation}\label{caldeiraleggett} 
m \ddot{x} \,+\, \eta \dot{x}\,+\, {\partial V\over \partial x}\, = \, 
F_{ext} \, . 
\end{equation}
The potential $V(x)$ and friction coefficient $\eta$ were regarded as 
experimentally determined quantities. 
Dissipative tunneling was later investigated by Ford {\it et al}. 
\cite{ford}, using a quantum Langevin approach.
The effect of dissipation has been studied in literature 
along two different lines: phenomenological approaches and microscopic 
formulations where dissipation comes about due to the coupling of the system 
to a heat bath of infinitely many degrees of freedom.    
In the present work we restrict ourselves to the former type of 
approaches. We study the effects of dissipation in a
solid state junction with an electron traversing through a 
metal - insulator - metal surface, within a model which is 
conceptually similar to that in \cite{caldeira} 
and consider a frictional force 
of the type $F(x) = \eta v(x)$, where, $v(x)$ is the effective velocity 
defined above (however, now with dissipation)  
and $\eta$ is the coefficient of friction. Such a model 
was proposed in \cite{buetikprl} and often used to represent the dissipation 
due to a damping force. Though in the microscopic treatments, dissipation is 
included by considering the entanglement of the system and the 
environment in a time dependent quantum 
mechanical approach \cite{qmdisip}, we restrict here to the 
simpler approach involving 
a frictional coefficient due to the following reasons: (i) the objective of the present 
work is to explore the behaviour and subtleties of the dwell 
times and traversal times 
which are concepts involving stationary wave functions. 
These concepts give a measure of 
time intervals without refering to the parametric time $t$ 
appearing for example in the time dependent 
Schr\"odinger equation. (ii) The question which we wish to address 
in the present work is 
``how does such a stationary time concept get affected by 
dissipation of the particle energy during tunneling"? and not a more global 
one of how the quantum tunneling is affected by the interaction 
with the environment which produces the dissipation. 
Such a semiclassical approach has also been used earlier 
in \cite{buetikprl,ranfagni2,bhataroyadmp}. We also refer the reader to 
\cite{pimpale} for a review on the various approaches.  

The introduction of the frictional force, $\eta \dot{x}$, is not entirely
arbitrary but can rather be derived under certain approximations from the complete
picture of a system coupled with an environment \cite{ingold,weissetc}.
The author in \cite{ingold} for example, 
starts with a model for the dissipative quantum system by
considering a Hamiltonian with three contributions coming from the system degrees
of freedom, the environment and the coupling between them. 
The system Hamiltonian models a particle of mass $m$ 
moving in a potential $V$ and the environment Hamiltonian
describes a collection of harmonic oscillators. After some discussions and algebra, the
author arrives at an equation of motion given by,
\begin{equation}\label{ingold} 
m \ddot{x}\, +\, m \displaystyle\int_0^t\, ds\, \gamma(t-s)\, 
\dot{x}(s)\, +\, {\partial V\over \partial x} \, =\, F_{ext}(t),
\end{equation} 
where $\gamma(t)$ is a damping kernel
which can be expressed in terms of the spectral density, $J(\omega)$,
of bath oscillators as,
\begin{equation}\label{dampkernel} 
\gamma(t) = {2 \over m} \displaystyle\int_0^{\infty}\, {d\omega \over \pi} \, 
{J(\omega) \over \omega} \, \cos{(\omega t)}\, .
\end{equation}
The most frequently used spectral density, $J(\omega) = m \eta \omega$, associated
with the so called ``Ohmic damping", then leads us back to Eq. (\ref{caldeiraleggett})  
of Caldeira and Leggett, mentioned above.

For a particle with energy $E$, 
tunneling a rectangular barrier of width $L$ and 
height $V_0$, the amount of energy 
lost while traversing a distance $x$ can be written as, 
$\Delta E(x) = \eta \int_0^x \,v(x^{\prime}) dx^{\prime} = 
 \eta [2(V_0 - E)/m]^{1/2} \,x$. This implies that at every $x$ inside the 
barrier, the energy of the particle is modified from $E \to E^{\prime} 
= E - \Delta E(x)$ and 
$\kappa_0 \,=\,{\sqrt{ 2 m \,(V_0\,-\,E)} / \hbar}$ in turn is modified to 
$\kappa (x) \,=\,{\sqrt{ 2 m \,(V_0\,-\,E \,+\,
\Delta E(x))} / \hbar}$. 
The maximum energy lost, $\Delta E (L)$, however, cannot exceed the energy of 
the tunneling particle, i.e., $\Delta E(L) \le E$. This puts a limit on the allowed 
value of the friction coefficient and we get, 
$\Delta E(L) = \eta_{max}\, [2(V_0 - E)/m]^{1/2} \,L = E$, leading to, 
\begin{equation}\label{etamax} 
\eta_{max} = {V_0 \over L} \, \biggl ( {E\over V_0} \biggr ) \, 
\Bigg[ {m \over 2 V_0 \biggl( 1 - {E\over V_0} \biggr)} \Bigg]^{1/2} \, . 
\end{equation}
In Fig. 2, we show the fraction of energy lost as a function of the energy of the 
tunneling particle for a couple of values of $\eta$ which are close to those 
determined in a realistic tunneling of an electron in a solid state junction 
(to be discussed in the next section). 
\begin{figure}[ht]
\includegraphics[width=9cm,height=9cm]{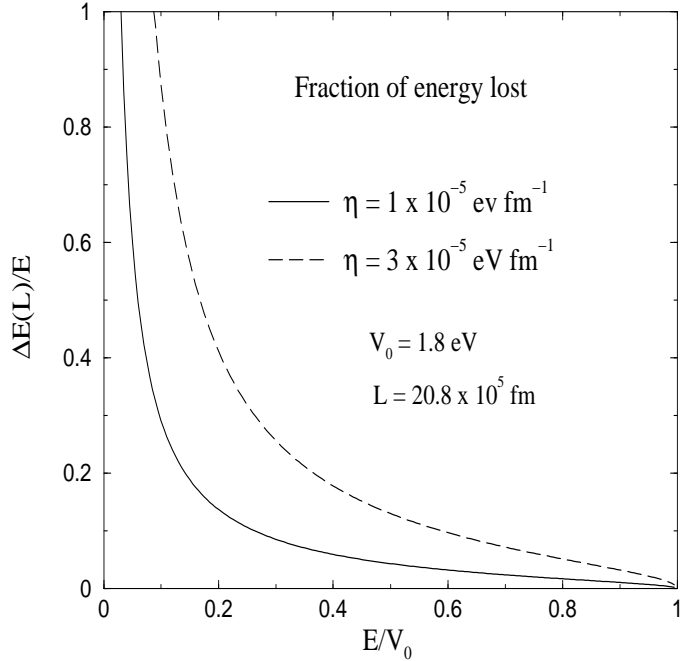}
\caption{\label{fig:eps2}
Fraction of energy lost as a function of energy of the tunneling particle. } 
\end{figure}

\subsection{Average dwell time and traversal time with dissipation} 
Defining, $V_0 + \Delta E(x) = V(x)$, gives 
$\kappa (x) \,=\,{\sqrt{ 2 \,m\,(V(x)\,-\,E)} / \hbar}$ and 
the dwell and traversal times with dissipation 
for a rectangular barrier can be evaluated within the JWKB approximation, using 
Eqs (\ref{dwellav}) and (\ref{traver}) for tunneling of a particle with 
energy $E$ through a potential barrier given by $V(x)$. The calculation is easily performed 
numerically. However, to get some insight into the relations, let us 
consider the tunneling of particles with a small amount of
dissipation, such that, 
$\Delta E(x) \ll V_0 - E$ and hence $\hbar \kappa (x)$ can be expressed 
approximately as, $\hbar \kappa (x) = \hbar \kappa_0 + \eta\, x$. 
Assuming further that, $\eta L \ll \hbar \kappa_0$, the average dwell time 
reduces to a rather simple expression given by (see appendix for details)  
\begin{equation}\label{dwelldisip}
\tilde{\tau}_D = {m \over 2 \, \hbar \kappa_0^2} \, \biggl [ \, 1 \, -\, 
\exp^{-2\,\kappa_0\, L\, -\, \eta\, L^2/\hbar} \, 
\biggl ( \, 1 - {2\eta L\over \hbar \kappa_0} \, \biggr) \, 
\biggr ].
\end{equation}
The transmission coefficient with dissipation can similarly be shown 
to be 
\begin{equation}\label{transdisip} 
\tilde{\cal T}(E) = e^{-\eta L^2/\hbar}\,e^{-2 \kappa_0\, L}\, , 
\end{equation}    
where $e^{-2 \kappa_0\, L}$ is the standard transmission coefficient in 
the JWKB approximation. The traversal time with dissipation, as shown in \cite{ranfagni} 
reduces to
\begin{equation}\label{travdisip} 
\tilde{\tau}_{tr} = {m \over \eta } \, \ln \biggl ( \, 1 \,+\, {L \over a}\, \biggr )  \, , 
\end{equation} 
with $a^2 = 2 m (V_0 - E) / \eta^2$, for $E < V_0$ as shown in \cite{ranfagni}. 
In order to ensure the correct behaviour of the dwell time 
close to the bottom of 
the barrier, the dwell time in Eq. (\ref{dwelldisip}) is multiplied by the 
factor $N^2$ suggested in the 
previous section on the improved JWKB expressions. 
\begin{figure}[ht]
\includegraphics[width=12cm,height=12cm]{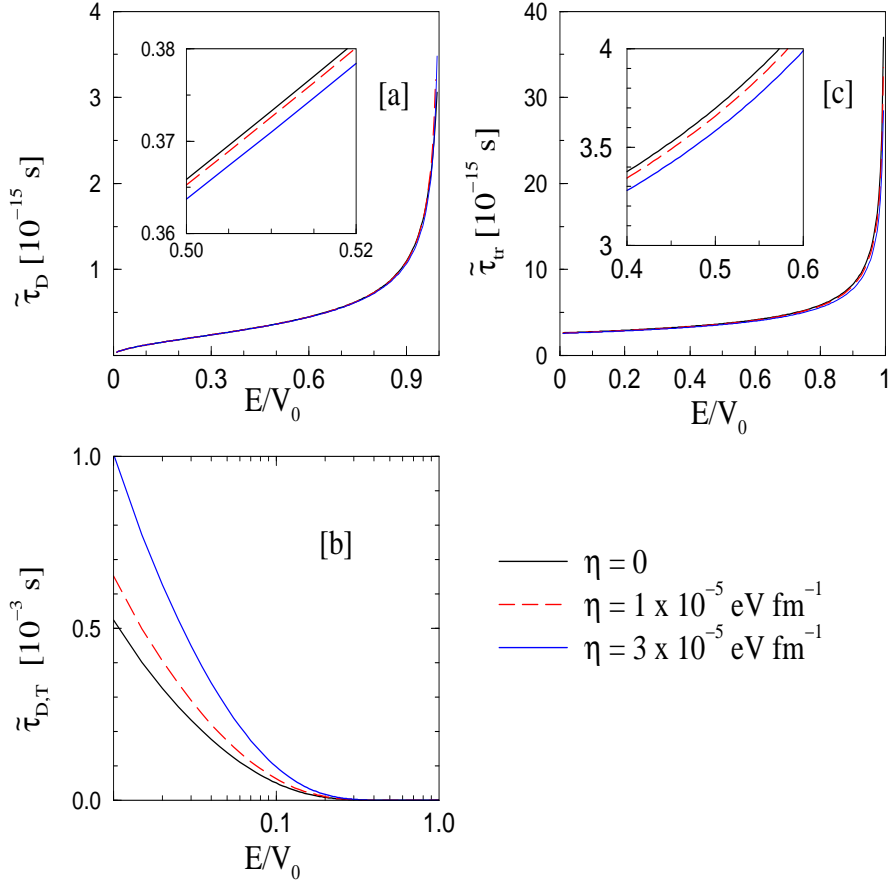}
\caption{\label{fig:eps3}
Average dwell time, $\tilde{\tau}_D$ [a], transmission dwell time, 
$\tilde{\tau}_{D,T}$ 
[b] and traversal time, $\tilde{\tau}_{tr}$ [c] for 
different amounts of dissipation in tunneling through a 
rectangular barrier of height $V_0$=1.8 eV and length 
$L = 20.8 \textrm{\AA}$.}
\end{figure}
 
\subsection{Reflection, Transmission and Average dwell time}
The definition of an average dwell time, $\tau_D$, is the time spent in a 
region, say, 
$(x_1, x_2)$ regardless of the fact if the particle escaped by reflection or
transmission. This $\tau_D = {\int_{x_1}^{x_2}\,|\Psi(x)|^2\,dx / j}\,$ 
is defined \cite{nusen} as the number density divided 
by the incident flux, namely, $j = \hbar \,k_0 \,/\mu$ 
(with $k_0 = \sqrt{2 \mu E} / \hbar$) for a free particle.
However, one can also define
transmission and reflection dwell times for the particular cases when the
particle, after dwelling in a region, escaped either by transmission or reflection. 
The flux $j$ in these cases would get replaced by the transmitted or
reflected fluxes,
$j_T = \hbar \, k_0 {\cal T}/ \mu$ and $j_R = \hbar \, k_0 {\cal R} / \mu$
\cite{gotoiwamoto} respectively. 
One would then obtain \cite{gotoiwamoto},
\begin{equation}\label{reciprocal} 
{1 \over \tau_D} \, =\, {{\cal T} \over \tau_D} \, + \, {{\cal R} 
\over \tau_D} \, 
= \, {1 \over \tau_{D,T}} \, + \, {1 \over \tau_{D,R}}
\end{equation}
where ${\cal T}$ and ${\cal R}$ are the transmission and reflection coefficients
(with ${\cal T} \, +\, {\cal R}\,=\, 1$ due to conservation of probability) and
$\tau_{D,T} = \int |\Psi|^2 dx / j_T$ and
$\tau_{D,R}  = \int |\Psi|^2 dx / j_R$,
define the transmission and reflection dwell times
respectively. In Ref. \cite{weEPL}, it was shown that it is the transmission dwell 
time, $\tau_{D,T}$ which can be attributed the physical meaning of the lifetime 
of a decaying nucleus. Within a semiclassical model for alpha particle tunneling, 
$\tau_{D,T}$ was shown to reproduce the half-lives of heavy nuclei.
Defining a transmission dwell time 
in the case of dissipative tunneling in the same manner as above, we 
can write, $\tilde{\tau}_{D,T} (E) = \tilde{\tau}_D(E)/\tilde{{\cal T}}(E)$ 
where 
$\tilde{\tau}_D(E)$ and $\tilde{{\cal T}}(E)$ as defined in 
(\ref{dwelldisip}) and (\ref{transdisip}) are the average dwell time and 
transmission coefficient with dissipation. It is also worth mentioning that 
in \cite{gotoiwamoto} it was shown that whereas the average dwell time 
saturates with increasing widths of a tunneling barrier (leading to 
speculations of superluminal propagation), the transmission dwell time 
does not. The latter was shown to simply increase with barrier width. 
Due to the fact that the transmission coefficient 
${\cal T}$ is usually much smaller
than ${\cal R}$, we have the average, $\tau_D \approx \tau_{D,R}$, i.e., the
average dwell time is similar to the reflection dwell time and
$\tau_{D,T} \gg \tau_{D,R}$. In the presence of
dissipation, there is an energy loss and
the transmission coefficient (calculated at a lower energy, $E - \Delta E$)
is reduced and hence
$\tilde{\tau}_{D,T} = \tilde{\tau}_D / \tilde{{\cal T}}$ 
increases. The reflection coefficient however,
increases and hence the reflection dwell time $\tau_{D,R}$ and hence the
average dwell time decrease.
Thus, in what follows, we shall see that another paradoxical 
situation which arises from 
the calculation of the average dwell time is resolved if one rather 
studies the behaviour of the transmission dwell time.

In Fig. 3a, we show the average dwell time with and without dissipation for 
a typical example of electron tunneling through a solid state junction. 
The details of the dissipation constant and its values will be discussed 
in a later section. In the inset in Fig. 3a, we can see that the dwell 
time with dissipation is less than that without dissipation. 
The traversal time with dissipation (see Fig. 3c) 
displays a similar behaviour too. 
This 
counter-intuitive result is however reversed when we calculate the 
transmission dwell time and a particle losing energy is indeed seen 
to spend a greater amount of time in the barrier, in Fig. 3b. This 
observation once again confirms the physical nature of the transmission
dwell time. 
Apart from this, we have also seen in Fig. 2 that the 
fraction of energy lost in the barrier decreases with increasing 
energy. As a result, the transmission dwell time, $\tilde{\tau}_{D,T}$, with 
dissipation 
can be seen in Fig. 3b to be similar to $\tau_{D,T}$ without dissipation for energies 
approaching the top of the barrier.

\section{Temperature dependent dissipation in Al$_2$O$_3$}
In an earlier paper \cite{weAPL} involving two of the authors of the present 
work, measurements of 
the current-voltage (I-V) characteristics of an Al/Al$_2$O$_3$/Al junction 
at temperatures from 3.5 to 300 K were reported. These data were 
used in order to extract the temperature dependence of 
characteristic quantities such as the barrier height and the average dwell 
time in the tunneling of an electron through the Al/Al$_2$O$_3$/Al junction. 
The rectangular barrier height $V_0(T)$ was found to 
decrease for temperatures increasing from 3.5 to 300 K. 
In what follows, we shall try to relate the temperature dependence of 
$V_0(T)$ found in \cite{weAPL}  
to the dissipation phenomenon and see if a correlation emerges.
In the model of the present work, 
the problem of dissipative tunneling through a rectangular barrier 
of fixed height $V_0$ gets modified to that of tunneling 
of a particle with energy, $E$, through 
an effective potential 
$V(x) = V_0 + \Delta E(x)$.  
If one starts with the assumption that 
the amount of energy dissipated in tunneling could change with temperature,
i.e., $\Delta E(x) \to \Delta E(x,T)$,  
the effective potential can be expressed as,
\begin{equation}\label{pottemp}
V(x,T) = V_0 \, +\, \eta(T) v_0 x\, , 
\end{equation}
where, $\Delta E(x,T) = \eta(T) v_0 x$ and the temperature dependence of 
the potential barrier is contained in the coefficient of friction, $\eta$. 
The transmission coefficient for a particle tunneling through the potential 
$V(x,T)$ is given by (\ref{transdisip}) with the difference that $\eta \to 
\eta(T)$. This means that the I-V data 
in \cite{weAPL} which were fitted using the commonly used Simmons' JWKB 
formula \cite{simmon},  
$J(V) =  \int_0^{E_m}\, {\cal T}(E) \xi(V) \, dE$, 
with a transmission coefficient ${\cal T}(E)$, 
can now be fitted using the transmission coefficient with dissipation, 
namely, 
$\tilde{{\cal T}}(E)$ instead of ${\cal T}(E)$. Since $\tilde{{\cal T}}(E)$  
in the JWKB approximation 
can be approximated (as shown in the appendix) for small amounts of energy 
dissipation by, 
$\tilde{{\cal T}}(E) = e^{-\eta(T) L^2/\hbar} {\cal T}(E)$, the current 
density $J(V)$ in 
Simmons' model gets modified to, 
\begin{equation}
\tilde{J}(V) =  e^{-\eta(T) L^2/\hbar} \,\int_0^{E_m}\, {\cal T}(E) \xi(V) \, dE\, , 
\end{equation}
with the factor $\xi(V)$ as given in \cite{simmon}.  
For a detailed derivation 
of the I-V relations (in the absence of dissipation) for low, 
intermediate and high voltages  
we refer the reader to the seminal papers of Simmons \cite{simmon} 
and proceed here by noting that in case of dissipative tunneling, 
$J(V)$ in \cite{simmon} simply gets modified 
to $\tilde{J}(V) = e^{-\eta(T) L^2/\hbar} \,J(V)$.  
Performing new fits to the I-V data with fixed barrier height, $V_0$, 
and fixed width, $L$, but 
the frictional coefficient $\eta$ as a free parameter which can vary with temperature, we find that the best fit is obtained 
for a constant barrier width and height of $L = 20.8 \textrm{\AA}$ and 
$V_0 = 1.799$ eV respectively. 
The latter is indeed the height of the barrier found 
in \cite{weAPL} at 300 K. 
In Fig. 4a, we present the fitted values of $\eta$ as a 
function of temperature. 
\begin{figure}[h]
\includegraphics[width=14cm,height=7cm]{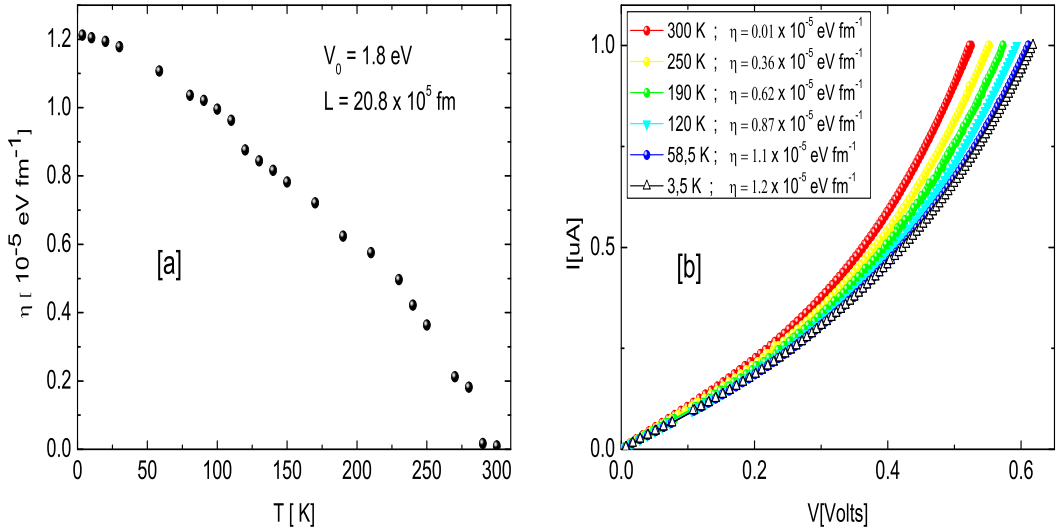}
\caption{\label{fig:eps4} 
Dissipation coefficient, $\eta$ as a function of temperature [a] and 
the I-V characteristics (reported in \cite{weAPL} and re-fitted 
here as explained in the text) at different temperatures [b] for 
electrons tunneling through a solid state junction of Al/Al$_2$O$_3$/Al. }
\end{figure}
We see a reduction in $\eta$ or a decrease in energy loss with 
increasing temperature. This behaviour is consistent with the results 
in \cite{weAPL}, 
where performing a fit without taking dissipation into account, we found that the 
barrier height, $V_0(T)$, decreases with increasing temperature. 
A smaller $V_0$ for a 
given energy $E$ corresponds to the tunneling of the particle closer to the 
top of the barrier where the energy loss is little (see Fig. 2). 
It is interesting to note that the same I-V data 
(see Fig. 4b) can be fitted either with a model 
including dissipation where the coefficient of friction, $\eta(T)$, 
decreases with temperature for a fixed barrier height $V_0$ or with 
$\eta =0$ and a 
temperature dependent barrier height $V_0(T)$. 
Thus, in a model which includes dissipation explicitly, 
the effect of the temperature dependence of the barrier height is 
reflected in a temperature dependence of the friction coefficient. 

\section{Summary} 
Dissipative tunneling of a particle such as an electron through a rectangular 
barrier has been investigated using the semiclassical JWKB approximation. 
Though the choice of an electron tunneling a rectangular barrier 
has been made in view of the 
analysis of the current-voltage data \cite{weAPL} which is re-analysed here 
with dissipation, the characteristics of the dwell and traversal times found 
here should hold in general. 
The average dwell time of a particle tunneling a barrier 
(i.e., the time spent in a region irrespective of the 
fact if the particle got reflected or transmitted after residing in the 
barrier) is reduced in dissipative tunneling. 
This counter-intuitive result is however 
reversed if one evaluates the transmission dwell time and 
dissipation of energy in the barrier is found to delay the tunneling 
of the transmitted particles. Apart from this, the percentage 
energy loss in tunneling is found to be maximum near the 
base of the barrier and decreases for energies approaching the top of the 
barrier. 

Based on the realistic example of an electron tunneling through a 
solid state junction, we find that the traversal time 
(as defined in (\ref{traver})) is an order of magnitude larger than the 
average dwell time spent by the electron in the barrier. Given the reciprocal 
relation (Eq. (\ref{reciprocal})) for the average, transmission and reflected 
dwell times, the transmission dwell time comes out to be orders of 
magnitude larger than the average dwell time (shown in Fig. 3). 
Fits to the current voltage (I-V) data in the Al/Al$_2$O$_3$/Al 
junction for temperatures ranging between 3.5 K and 300 K, using the 
dissipation model of the present work display a decreasing dissipation 
(frictional) coefficient as a function of increasing temperature. 
The method may prove useful to study dissipative effects in junctions with 
other materials too. 

\acknowledgments
One of the authors (E. J. P.) wishes to thank ``Convocatoria Programas 2012", 
Vicerrector{\'i}a de Investigaciones, ``Proyecto Semilla", 
Facultad de Ciencias and ``Convocatoria para la Financiaci{\'o}n de 
Inversiones en Equipos de Laboratorio", 
Departamento de F{\'i}sica of Universidad de los Andes (Bogot{\'a}, Colombia) 
for financial support.

\appendix*
\section{Dwell time with dissipation}
Noticing that the effects of dissipation are mostly relevant at the base of the 
barrier, an expression for the dwell time with dissipation can be derived 
analytically within some reasonable approximations. We find that the difference 
between the numerical results presented in this work and those evaluated 
from the analytical expression as in Eq. (\ref{dwelldisip}) is negligible. 
Here we give a brief derivation of Eq. (\ref{dwelldisip}) and the 
transmission coefficient with dissipation in the JWKB.

The effective velocity for a particle tunneling a rectangular barrier of height $V_0$ 
can be written as $v_0 = \hbar \kappa_0/m$, with 
\begin{eqnarray}
v_0=\frac{\hbar k}{m}=\sqrt[]{\frac{2}{m}\left(V_0-E\right)} \qquad 
\kappa_0=\frac{\sqrt{2m\left(V_0-E\right)} }{\hbar}\nonumber
\end{eqnarray}
Including dissipation in the form of a frictional force, the energy loss can be given 
as a function of distance as, $
\Delta E(x)=\eta \int_0^x v_0 dx^\prime \,=\,\eta \sqrt{2(V_0-E)/m} \,x$.  
This means that $\kappa_0$ gets modified to $\kappa(x)$ 

\begin{eqnarray}\label{kappas} 
\kappa_0\to \kappa(x)&=&\frac{\sqrt{2m(V_0-E^\prime)}}{\hbar}\nonumber\\
&=&\frac{\sqrt{2m\left[V_0-E+\Delta E(x)\right]}}{\hbar}\nonumber\\
&=&\frac{\sqrt{2m(V_0-E)}}{\hbar}\left(1+\frac{\eta}{\hbar k }x\right) \quad;\quad \Delta E \ll V_0-E\nonumber\\
\kappa(x) &=& \kappa_0\left(1+\frac{\eta}{\hbar k }x\right)
\end{eqnarray}
Replacing this expression of $\kappa(x)$ into 
Eq. (\ref{dwellav}), the average dwell time with dissipation is given as, 
\begin{eqnarray}
\tilde{\tau}_D&=&\frac{\mu}{\hbar}\int_0^L \frac{1}{k} \cdot exp\left(-2kx-\frac{\eta x^2}{\hbar}\right) dx -\frac{\eta \mu}{ \hbar^2 k^2}\int_0^L x \cdot exp\left(  -2kx-\frac{\eta x^2}{\hbar} \right)dx
\end{eqnarray}
The above integral can be solved analytically using the following formulae: 
\begin{eqnarray}
\int exp\left(  -ax- b x^2 \right) dx &=&\frac{\sqrt{\pi} \cdot e^{a^2/4b} erf\left(\frac{a+2bx}{2\sqrt{b}}\right)  }{2\sqrt{b}}\nonumber\\
\int x\cdot exp\left(  -ax- b x^2 \right) dx &=&-\frac{a\sqrt{\pi}  \cdot e^{a^2/4b} erf\left(\frac{a+2bx}{2\sqrt{b}}\right)}{4b^{3/2}} - \frac{e^{-x(a+bx)}}{2b}\nonumber
\end{eqnarray}
After some lengthy but straightforward algebra, the expression in terms of the 
$erf$ functions can be further simplied by using the expansion, 
for $ x\gg 1 $, 
\begin{eqnarray}
erf(x) = 1- \frac{e^{-x^2}}{\sqrt{\pi} x}\left[  1- \frac{1}{2x^2}- \cdots   
\right]\,,
\end{eqnarray}
with the assumption that $\eta L \ll \hbar \kappa_0$. This leads us to the dwell time 
given in Eq. (\ref{dwelldisip}). The transmission coefficient in the JWKB, namely, 
$T = \exp{ (-2 \, \int_0^L\, \kappa(x)\, dx)}$, with 
$\kappa(x)$ defined as in (\ref{kappas}) reduces to, 
\begin{eqnarray}
\tilde{T}(E) &=& \exp{(-2 \int_0^L\,[\kappa_0 + \eta x/\hbar]\, dx) }\nonumber \\
&=& \exp{[-\eta L^2/\hbar]}\, {\cal T}(E)\, , 
\end{eqnarray}
where ${\cal T}(E) = \exp{(-2 \kappa_0 L)}$, is the transmission coefficient 
without dissipation.

\end{document}